\documentclass[%
 reprint,
superscriptaddress,
 amsmath,amssymb,
 aps,
]{revtex4-2}

\usepackage[nice]{nicefrac}
\usepackage{graphicx}
\usepackage{dcolumn}
\usepackage{multirow}
\usepackage{bm}
\usepackage{color}

\begin{document}

\preprint{APS/123-QED}

\title{Super narrow peaks in excitation spectrum of alkali spin polarization:\\non-adiabatic case of spin dynamics} 

\author{E. N. Popov}
 \email{enp-tion@yandex.ru}
 \affiliation{
 Laboratory of Quantum Processes and Measurements, ITMO University,\\199034, 3b Kadetskaya Line, Saint-Petersburg, Russia
}
\author{A. A. Gaidash}
\author{A. V. Kozubov}
\affiliation{
 Laboratory of Quantum Processes and Measurements, ITMO University,\\199034, 3b Kadetskaya Line, Saint-Petersburg, Russia
}
\affiliation{Department of Mathematical Methods for Quantum
  Technologies, Steklov Mathematical Institute of Russian Academy of
  Sciences, 119991, 8 Gubkina St, Moscow, Russia} 


\author{S. P. Voskoboynikov}
\affiliation{
 Higher School of Software Engineering, Peter the Great St.Petersburg Polytechnic University,\\195251, 29 Polytechnicheskaya, Saint-Petersburg, Russia
}%

\date{\today}

\begin{abstract}

We theoretically describe the phenomenon of non-adiabatic spin dynamics, which occurs in a gas cell filled by alkali vapor in presence of a strong alternating magnetic field and pump light. Steep increase of the spin polarization occurs if frequency of the magnetic field is equal to the certain value. Although, the observable effect relies on the periodic field that consists of two perpendicular components defined by harmonics with the same amplitudes and different frequencies. Considered spin effect cannot be explained by a resonance, because the own Larmor frequency of spin precession is absent without a constant component of magnetic field. Moreover, there are some clearly visible peaks in the excitation spectrum of spin polarization, and they are super narrow in comparison to relaxation rate. Detailed analysis according to proposed quantum model results in the reasoning of the effect via qualitative properties of non-adiabatic dynamics of atomic spin. 


\end{abstract}

\maketitle


\section{\label{sec:level1}INTRODUCTION}

The atomic spin dynamics in a gas cell has been studied for over half of the century. The main advantage of that physical model is the large lifetime of non-excited spin states in comparison to crystals and liquids, since any atom moves as a free particle between collisions \cite{minute:decay,balabas:10,advances:decay}. Moreover, in a mixture consisting of an alkali vapor and an inert buffer gas, only a small percentage of collisions results in spin reversal. These factors make it possible to observe some evident magnetic effects in an alkali vapor, which are implemented in precise measurements for metrology and navigation \cite{epr:book1,epr:book2}, i.e. magnetometers \cite{pomerantsev,farr,vershovsky,nature:magnetometer,bison,budker:kimball,alexandrov:vershovsky,zhivun}, gyroscopes \cite{review,Donley,Vershovsky:gyro,konrack,larsen:gyro,gyro:signal,gyro:new1,gyro:new2} and frequency standards \cite{IEEEst,radioFr}. A rapid development in the study of a gas cell with spin dynamics is induced by the implementation of optical approaches to scanning and excitation of a spin polarization \cite{happer:article,happer:book,kubo,aleksandrov,happer:edu,suter,skolnik,kozlov,talker, orientation,Kastler:63,nobel}.

In an alkali vapor, one of the most explored and utilized in the practice magnetic effects is the electron paramagnetic resonance (EPR) \cite{zavoisky,first:res,bell:bloom}. A necessary condition for the EPR observation is the presence of a constant magnetic field, which defines the own Larmor frequency of the alkali spin precession. If the frequency of periodic perturbation of the spin system is close to or equal to the own Larmor frequency, the EPR is arising as oscillations of spin polarization. The width of the EPR spectral line is determined by a rate of population mixing among alkali Zeeman sublevels. The latter is the most important parameter of the system, because the precision of different gas cell sensors depends on lifetime of spin states \cite{cesium:transitions,sodium:relaxation,rubidium:relaxation,rubidium:kr,bouchiat,quad:relaxation,franz:franz,coating:relaxation,thesis:breault, cells}.

In this work, we theoretically explore a phenomenon in alkali vapor: its spin polarization steeply increases near a certain frequency of an external magnetic field in the presence of a circularly-polarized pump light. To observe the effect, the following properties of the magnetic field should be satisfied:

\begin{itemize}
\item
The external magnetic field is periodic with the period that is much less than the relaxation time of a spin polarization;
\item
The averaged over a period external magnetic field is equal to zero, thus the constant component of the magnetic field in a gas cell with alkali vapor is absent;
\item
Both the magnitude and direction of the magnetic field vary;
\item
Zeeman shift of sublevels with non-zero angular momentum projection is much larger than broadening of these sublevels due to population mixing to equilibrium state.
\end{itemize}

Listed above properties regard to non-adiabatic dynamics, defined by rapid variation of an alkali spin polarization in comparison to its relaxation. Withal, magnetic field should be periodic to analyze the steady dynamics through its constant characteristics. It is essential, that the origin of the effect differs from EPR induced by an externally applied periodic force. Though, to exclude an EPR we need to disable a trend of Larmor spin precession, therefore a constant component of magnetic field should be absent. Hereafter in the paper, we use a term {\it the spin effect} to refer to the phenomenon under consideration.

The listed above properties may belong to a magnetic field defined by various temporal dependencies. Here, we consider the special case that is sufficient for exploring {\it the spin effect}:
\begin{gather}
\mathbf{B}(t)=B_0 \mathbf{l_z}\cos\left(\Omega t\right)+B_0 \mathbf{l_x} \cos\left(2\Omega t\right),\label{mag:field}\\
\gamma B_0 \gg \Gamma,\qquad \Omega \gg \Gamma,
\end{gather}
\noindent where $B_0$ is an amplitude of the magnetic field, $\mathbf{l_x}$ and $\mathbf{l_z}$ are orthonormal spatial vectors, $\Omega$ is a frequency of the magnetic field, $\gamma$ is gyromagnetic ratio of the oriented alkali atoms, $\Gamma$ is the relaxation rate of spin polarization.

Previously, various magnetic effects has been studied in condition of the non-adiabatic dynamics of oriented spins. There are some experimental and theoretical works, in which non-adiabatic dynamics is essential 
\cite{Applet,GAITAN1999152,recent:Budker,recent:spinint,recent:otherS,recent:response,recent:parametric,zero-field,sensor,doubleFr}. 
However, the cited papers do not focus on the non-adiabatic dynamics and it's properties. Moreover, to the best of our knowledge, the synchronization of alkali spin motion due to periodic non-adiabatic dynamics in presence of a strong alternating magnetic field was not analyzed in detail. We suggest exploring a spin effect induced by properties of non-adiabatic dynamics using a quantum model of alkali vapor.

\begin{figure}[b]
\includegraphics[width=8cm]{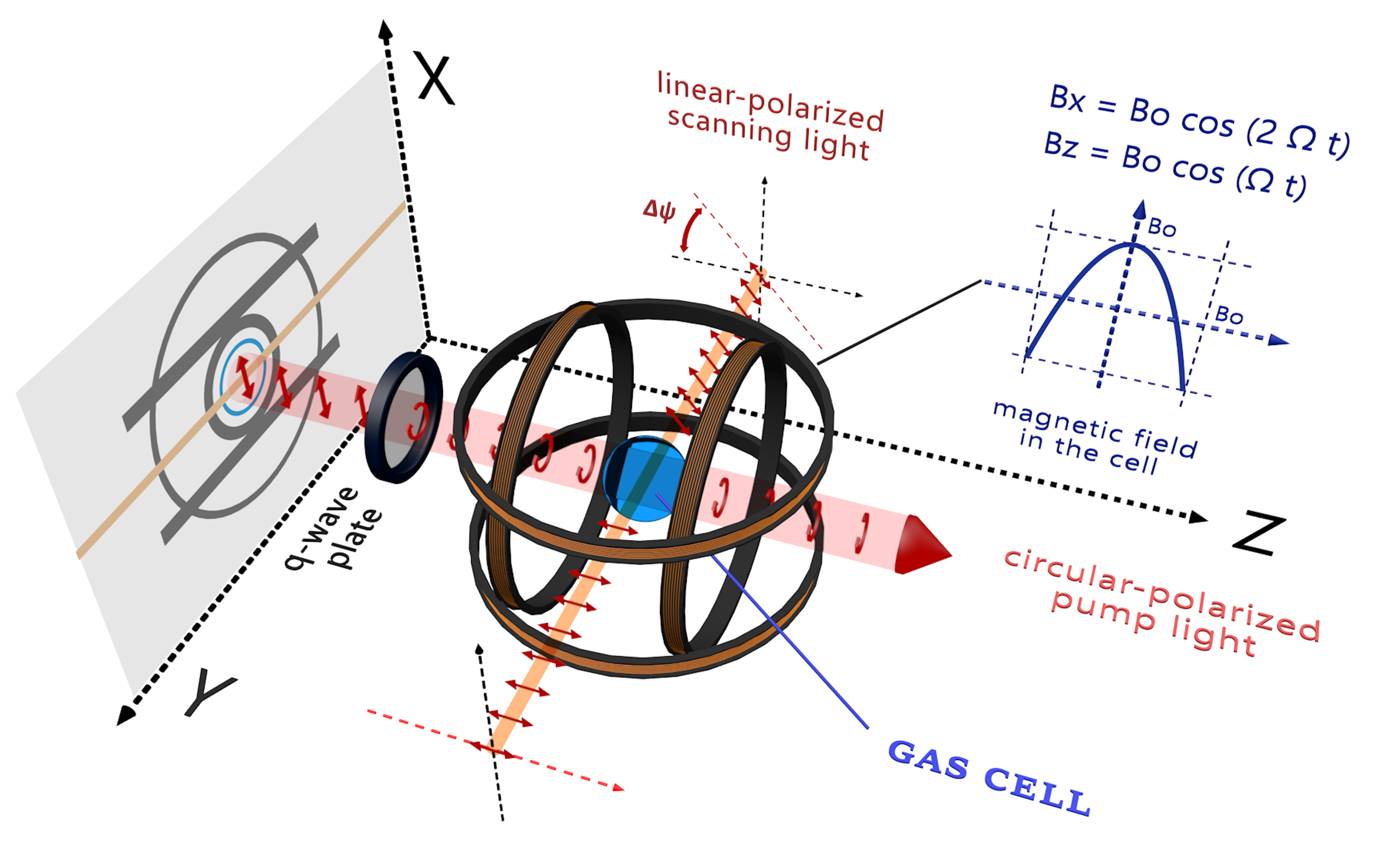}
\caption{\it \small The physical system for studying alkali spin dynamics in a gas cell. To create a magnetic field, the double Helmholtz coils located around the cell. A pump light propagates along the Z-axis and covers the cell homogeneously. A scanning light may be directed along any axis. The atomic spin projection onto the path of scanning light can be measured by the angle of scanning light polarization rotation $\Delta\psi$; they are proportional to each other}\label{pic:scheme}
\end{figure}

{\it The spin effect} cannot be explained in terms of spin resonance, since a constant magnetic field is absent. Parametric resonance \cite{parres} is not a suitable candidate for explanation of {\it the spin effect} as well, since there are no internal parameters of the atom-field system, which determine the own frequency of the resonance and make periodic oscillations. Though, we explain {\it the spin effect} by the qualitative properties of non-adiabatic dynamics of atomic spins in alkali vapor. Mathematically, the non-adiabatic case of atomic spin motion in the presence of a strong alternating magnetic field corresponds to a dynamic system with unpredictable behavior. Since the dynamics cannot be described as an explicit temporal dependence, the search for physical effects becomes a non-trivial task. Only via a numerical solution, we can explore the phase-space trajectory of a non-adiabatic dynamic system, that is considered in the paper.

The paper is organized as follows. In Sec.~\ref{sec_scheme} we describe an optical method of spin driving inside a gas cell with vapor of alkali. Also, the simplified scheme for an experiment is suggested. In Sec.~\ref{sec_equations} we propose a general quantum model of alkali spin dynamics in presence of a strong alternating magnetic field and relaxation. Calculations according to the model and resulting excitation spectrum of {\it the spin effect} are demonstrated in Sec.~\ref{sec_calculations}. In Sec.~\ref{sec_analysis} the explanation of {\it the spin effect} is provided. Sec.~\ref{sec_conclusion} concludes the article.

\section{Physical system under consideration}\label{sec_scheme}

In this section, we propose a principal optical scheme for {\it the spin effect} observation, see Fig.~\ref{pic:scheme}. Also, a structure of Rb-87 energy levels with excited transitions is shown in the Fig.~\ref{pic:levels}. The scheme consists of a gas cell with vapor of $^{87}$Rb and buffer gas, Helmholtz coils for generating the magnetic field Eq.~\eqref{mag:field}, a source of circularly polarized pump light (linearly polarized laser and quarter-wave plate) and source of scanning light (low-power linearly polarized laser). The pump light orients rubidium spins along the Z-axis, and the scanning light detects an expectation value of the atomic spin projection onto the path of the scanning light propagation. Note, the similar scheme of spin perturbation was considered in the works \cite{sensor} and \cite{doubleFr} in partially-adiabatic case.

\subsection{Gas cell}\label{opt_sch_a}

A gas cell contains a vapor of $\mathrm{^{87}Rb}$ and a mixture of inert gases. The gas mixture composes of diatomic nitrogen in order to reduce alkali fluorescence and some noble monatomic gases as a buffer. It is important that spin polarization can be formed by alkali atoms populating different hyperfine ground levels. Therefore, to define a certain group of non-excited alkali atoms by $\mathrm{Rb_{F=1}}$ and $\mathrm{Rb_{F=2}}$.

\begin{figure}[t]
\includegraphics[width=8cm]{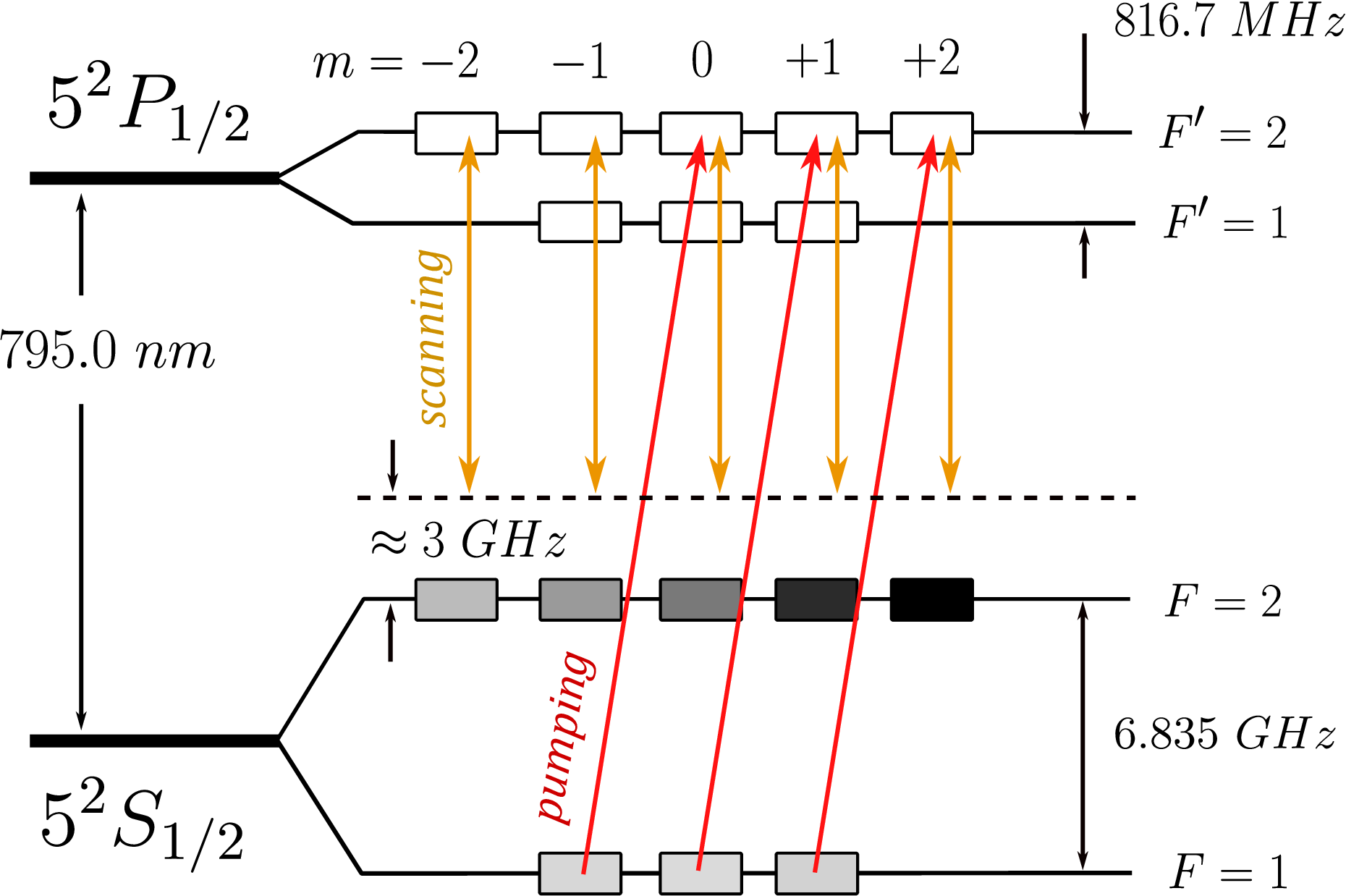}
\caption{\it \small D1 line of Rubidium 87. Rectangles denote Zeeman sublevels in the hyperfine structure. Degree of black filling shows its population: the darker, the higher population. Red arrows indicate atomic transitions \makebox[2.3 cm][s]{$F=1\leftrightarrow F'=2$} induced by circularly polarized pump light. Orange arrows indicate weak perturbation on transitions \makebox[2.3 cm][s]{$F=2\leftrightarrow F'=1$} and \makebox[2.3 cm][s]{$F=2\leftrightarrow F'=2$} excited by linearly polarized scanning light. Symbols $F$ and $F'$ denote total angular momentum of ground and excited levels respectively, $m$ is a projection of total angular momentum onto the path of the pump light propagation}\label{pic:levels}
\end{figure}

Temperature within the cell is about $80^\circ\,\mathrm{C}$. Under the conditions, the concentration of the alkali is 5--6 orders of magnitude less than the concentration of the buffer. Since an electron cloud of alkali has an enlarged radius and anisotropic form, multiple collisions between an excited $\mathrm{^{87}Rb}$ atom and $\mathrm{N}_2$ molecules lead to rapid non-radiative transitions from the upper to one of the two ground levels in the $\mathrm{D1}$~line (see Fig.~\ref{pic:levels}). Moreover, since the buffer environment freezes free motion of alkali, relaxation due to collisions with a cell's surface is significantly small (up to be considered absent). The latter properties of the gas cell are necessary for the inducing of non-adiabatic dynamics of atomic spins in an external magnetic field.

\subsection{Spin polarization pumping}\label{opt_sch_b}

As shown in the Fig.~\ref{pic:scheme}, a circularly polarized pump light propagates along the Z-axis. It orients alkali spins along the path of propagation. The frequency of a pump light is equal to the one of a transition between the levels with a total angular momentum \makebox[0.8 cm][s]{$F\!=\!1$} and \makebox[0.9 cm][s]{$F'\!=\!2$}. In the presence of an optical pump, alkali atoms populate the upper levels with one-sided change of the angular momentum, which is projected along the path of the light propagation. Therefore, alkali atoms accumulate non-zero angular momentum directed along the Z-axis. Recent known data about the D1~line of $\mathrm{^{87}Rb}$ can be found in \cite{steck}.

According to the optical scheme shown in the Fig.~\ref{pic:scheme}, spins of $\mathrm{Rb_{F=1}}$ and $\mathrm{Rb_{F=2}}$ are oriented through {\it depopulation} and {\it repopulation} effects, respectively. Both effects create the non-equilibrium population among Zeeman sublevels belonging to the ground levels of the D1~line. The nature of depopulation is a selective depletion of certain sublevels by a circularly polarized light. It should be noted, that repopulation process is more efficient than depopulation. As presented in \cite{scheme:re}, repopulation is based on the feature of collisional decay, that is a nucleus spin state is not destroyed during transition of an excited alkali atom from upper to ground levels. Therefore, spin polarization of $\mathrm{Rb_{F=2}}$ is produced by preserved alkali nucleus spin.

\subsection{Spin polarization scanning}\label{subsection:scanning:light}

Atomic ensemble with a spin polarization becomes circular birefringent due to inducing of optical anisotropy. The magnitude of the circular birefringence is determined by the expectation value of the atomic spin projection onto the path of light propagation. Since refraction indices for orthogonal circular light components are different, the linear polarization plane of the passing light rotates around the path of propagation. Therefore, alkali spin polarization dynamics can be measured via deviation of scanning light polarization as follows:
\begin{equation}\label{angular:proportional}
\langle\hat{F}_\alpha\rangle\propto\Delta\psi(t),
\end{equation}
\noindent where $\Delta\psi(t)$ is the angular rotation of the polarization plane shown in the Fig.~\ref{pic:scheme}, $\hat{F}_\alpha$ is the operator of the total angular momentum projection onto the $\alpha$-axis. By selection of the scanning light direction, one may measure any spin polarization component of the alkali vapor in the gas cell.

The spectral linewidth of the scanning light should be narrow enough for resolution of the alkali hyperfine structure. It is necessary for measurement of the spin polarization of alkali vapor only from $\mathrm{Rb_{F=2}}$. As shown in Fig.~\ref{pic:levels}, the scanning light frequency should be close to  \makebox[2.5 cm][s]{$F=2\leftrightarrow F'=2$} transition frequency. At the same time, it should be far from the frequency of \makebox[2.4 cm][s]{$F=1\leftrightarrow F'=2$} transition. If the latter conditions are satisfied, then $\mathrm{Rb_{F=2}}$ atoms affect the polarization of the scanning light significantly stronger than $\mathrm{Rb_{F=1}}$ atoms do. Withal, the light frequency should be detuned from the optical resonance in order to exclude redundant depletion of the ground level with $F=2$. Since the ground level with $F=1$ is broadened by a pump light, the spin polarization measurements from $\mathrm{Rb_{F=1}}$ are not advisable.

\section{Master equation}\label{sec_equations}

Here, we would like to propose the math model of alkali dynamics affected by a monochromatic light and an alternating magnetic field. To correctly describe {\it the spin effect}, we take into account common relaxation processes listed in the Table~\ref{table1}. The model is based on the semi-classical theory of atom-field interaction: alkali vapor is described by a density operator, and the fields are described by classical vectors dependent on time. The density operator is a $16\times 16$ matrix in the basis of non-perturbed Zeeman sublevels that belong to D1~line of $\mathrm{^{87}Rb}$.

\begin{table}[h]
\caption{\centering \it \small Relaxation processes in the gas cell, their origins and typical rates with corresponding mapping}\label{table1}
\begin{ruledtabular}
\begin{tabular}{ccc}
         Process& 
         Origin of the&
		Rate,\\ 
        description& 
        process&
	Mapping \\ 
        \colrule
        \\
        Mixing of ground&
        Alkali spin exchange& 
        \multirow{3}{*}
        {$\begin{array}{c}
        \Gamma\sim 10^{2}\:\text{Hz},
        \vspace{0.1cm}\\ 
        \hat{\rho}-\hat{\rho}_0
        \end{array}$} \\
        Zeeman sublevels&
        and atomic collision&
        \\
        population&
        with cell's walls& 
        \\
        \\
        Decay from excited&
        Non-ellastic collisions&
        \multirow{3}{*}
        {$\begin{array}{c}
        \delta_{dcy}\sim 10^{8}\:\text{Hz},
        \vspace{0.1cm}\\ 
        \mathcal{R}\left\{\hat\rho\right\}
        \end{array}$} \\
        to ground levels&
        between excited alkali&
        \\
        without fluorescence&
        and buffer atoms&
        \\
        \\
        Inhomogeneous&
        Velocities mixing to&
        \multirow{3}{*}
        {$\begin{array}{c}
        \delta_{mix}\sim 10^{9}\:\text{Hz},
        \vspace{0.1cm}\\
        \mathcal{M}\left\{\hat\rho\right\}
        \end{array}$} \\
        Doppler broadening&
        Maxwell-Boltzmann&
        \\
        of optical resonance&
        distribution&
        \\
        \\
        Decoherence of dipole&
        Alkali-buffer&
        \multirow{3}{*}
        {$\begin{array}{c}
        \delta_{dec}\sim 10^{10}\:\text{Hz},
        \vspace{0.1cm}\\
        \mathcal{D}\left\{\hat\rho\right\}
        \end{array}$} \\
        oscillations on&
        elastic collisions&
        \\
        optical transitions&
        without decay&
        \\
        \\
\end{tabular}
\end{ruledtabular}
\end{table}

The master equation for alkali density matrix, that describes the certain velocity group inside the Maxwell distribution, is as follows:
\begin{equation}\label{master:equation}
\begin{array}{l}
\displaystyle i\hbar\left(\frac{\partial\hat\rho}{\partial t}+k v_z\hat\rho\right)=\left[\hat H,\hat\rho\right]- \Gamma\left(\hat\rho-\hat\rho_0\right)-\vspace{0.2 cm}\\ 
\displaystyle 
-\delta_{mix}\mathcal{M}\left\{\hat\rho\right\}
-\delta_{dcy}\mathcal{R}\left\{\hat\rho\right\}
-\delta_{dec}\mathcal{D}\left\{\hat\rho\right\},
\end{array}
\end{equation}
\noindent where $\hat{H}$ is the Hamiltonian without relaxation, $k$ is the wavenumber, $v_z$ is the alkali velocity projection onto the path of the pump light, operator $\hat\rho_0$ corresponds to the state of thermodynamic equilibrium with mixed population among Zeeman sublevels. Operator $\hat{H}$ comprises the interaction of alkali with the pump light $\hat{V}_E$, the interaction of alkali with magnetic field $\hat{V}_B$ and unperturbed Hamiltonian $\hat{H}_0$:
\begin{gather}
\hat{H}=\hat{H}_0+\hat{V}_E+\hat{V}_B,\\
\hat{V}_E=-\left(\mathbf{\hat{d}\cdot E}\right), \qquad 
\hat{V}_B=\sum\limits_{n=1}^2 g_n\gamma_e \left(\mathbf{\hat{F}_n\cdot B}\right),\label{V:eq}\\
\mathbf{E}=\frac{\mathcal{E}}{2}\,\mathbf{l}_+ e^{-i\omega t}+c.c.,\qquad k=\frac{\omega}{c}.
\end{gather}
\noindent Here $\mathcal{E}$ is a constant amplitude of the pump light, $\mathbf{l}_+$ is the unit vector of the circular polarization, $\omega$ is the frequency of the pump light, the dipole operator $\mathbf{\hat{d}}$ describes all optical transitions between Zeeman sublevels in D1~line of $\mathrm{^{87}Rb}$ \cite{steck}, $\gamma_e$ is the electron gyromagnetic ratio, $g_n$ is the g-factor of the ground hyperfine levels, $\mathbf{\hat{F}}_n$ is the total angular momentum operator corresponding to atoms $\mathrm{Rb_{F=n}}$, and magnetic field $\mathbf{B}$ is defined by Eq.~\eqref{mag:field}.

Below we define mappings $\mathcal{M}$, $\mathcal{D}$ and $\mathcal{R}$ from the master equation~\eqref{master:equation}:
\begin{enumerate}
    \item The mapping $\mathcal{M}$ determines the transition of the dependence of density operator on alkali velocities to the uniform distribution:
\begin{equation}
\mathcal{M}\left\{\hat\rho\right\}=\hat\rho-\int\limits_{-\infty}^{\infty}\hat\rho\:\mu(v_z)\,\mathrm{d}v_z,
\end{equation}
\noindent where $\mu(v_z)$ is the Maxwell–Boltzmann distribution.

\item
The mapping $\mathcal{D}$ nullifies non-diagonal elements with an optical frequency of phase rotation:
\begin{equation}
\mathcal{D}\left\{\hat{\rho}\right\}=
-\hat{P}_{e}\hat{\rho}\hat{P}_{g}
-\hat{P}_{g}\hat{\rho}\hat{P}_{e},
\end{equation}
\noindent where $\hat{P}_{g}$ and $\hat{P}_{e}$ are projection operators to the space of ground hyperfine levels $F$ and excited hyperfine levels $F'$ in D1~line respectively, see Fig.~\ref{pic:levels}.

\item
The mapping $\mathcal{R}$ decomposes the full density matrix $\hat{\rho}$ to a tensor product of an electron $\hat{\rho}_{e}^{(0)}$ and nuclear $\hat{\rho}'_{n}$ density matrices:
\begin{gather}
\mathcal{R}\left\{\hat{\rho}\right\}=-\hat{P}_{e}\hat{\rho}\hat{P}_{e}+\hat{\rho}'_{n}\otimes\hat{\rho}_{e}^{(0)},\\
\hat{\rho}'_n=\sum\limits_{m=-\nicefrac{1}{2}}^{\nicefrac{1}{2}}
\langle{m;\,\scriptstyle 5^2P_{1/2}}\,|\,
\hat{\rho}
\,|\,m;\,{\scriptstyle 5^2P_{1/2}}\rangle,\label{nucleus_spin}\\
\hat{\rho}_{e}^{(0)}=\sum\limits_{m=-\nicefrac{1}{2}}^{\nicefrac{1}{2}}\frac{1}{2}
\,|\,m;\,{\scriptstyle 5^2S_{1/2}}\rangle
\langle m;\,{\scriptstyle 5^2S_{1/2}}\,|,
\end{gather}
\noindent where $|m;level\rangle$ is a spin-orbital state of an external electron with a projection $m$ of total angular momentum.
\end{enumerate}

\noindent Note, the electron density matrix reduces to the one of equilibrium state, however the nuclear density matrix does not. The partial trace in Eq.~\eqref{nucleus_spin} leads to a non-equilibrium distribution of population among Zeeman sublevels of $\mathrm{Rb_{F=2}}$. Due to this feature, alkali spin polarization during collision decay is preserved \cite{franz:franz, scheme:re}.

\begin{figure*}
\includegraphics[width=17cm]{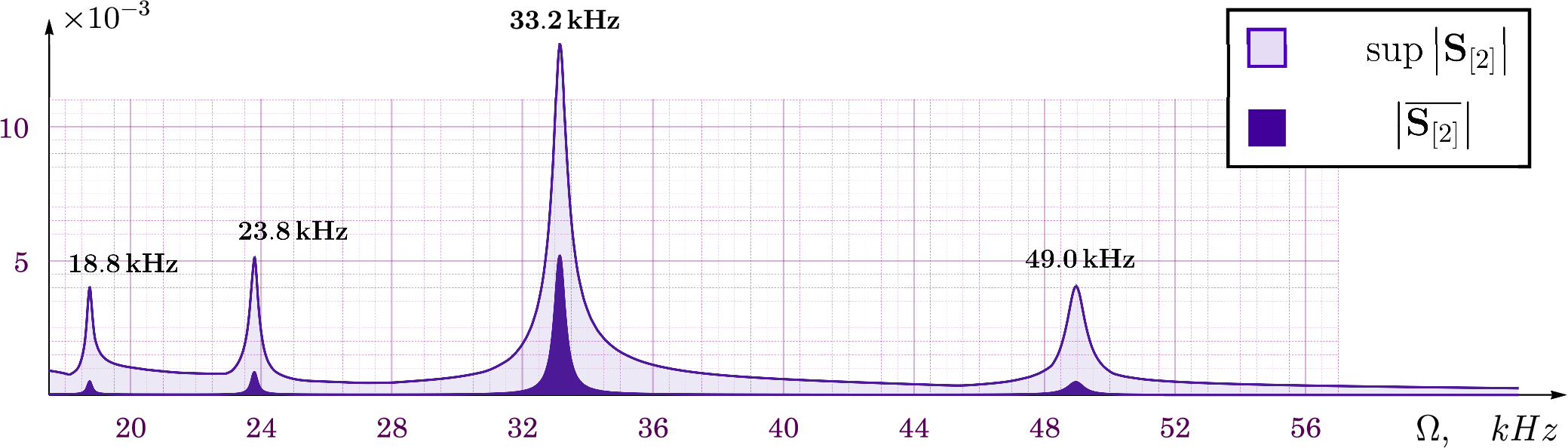}
\caption{\it \small Peaks in the $Rb_{F=2}$ spin polarization dependence on the frequency of periodic magnetic field determined by Eq.~\eqref{mag:field}; curves are defined by Eqs.~\eqref{conv:1} and~\eqref{conv:2}, correspondingly. Parameters for the evaluation are as follows: magnetic field amplitude $B_0=27\ \mu T$, the relaxation rate $\Gamma=1\ kHz$, the pump light amplitude $\mathcal{E} = 100\ V/m$, the temperature is $80^\circ \mathrm{C}$}\label{pic:freq:dep}
\end{figure*}

\section{Non-adiabatic spin dynamics evaluation}\label{sec_calculations}


In order to proceed with the numerical solution of the master equation~\eqref{master:equation}, let us prior define the vectors $\mathbf{S_{[1]}}$ and $\mathbf{S_{[2]}}$ for describing a spin polarization formed by alkali atoms $\mathrm{Rb_{F=1}}$ and $\mathrm{Rb_{F=2}}$ respectively:
\begin{gather}
\mathbf{S_{[n]}}=S_{[n],x}\mathbf{l_x}+S_{[n],y}\mathbf{l_y}+S_{[n],z}\mathbf{l_z},\label{S:n}\\[0.3cm]
S_{[n],\alpha}=
\mathrm{Tr}\left\{\hat\rho_n \hat\Sigma_{n,\alpha}\right\},\qquad\alpha\in\{x,y,z\},\label{S:na}\\
\hat\rho_n=\int\limits_{-\infty}^{\infty}\hat P_n\hat\rho\hat P_n \mu(v_z)\,\mathrm{d}v_z,\label{ro:n}
\end{gather}
\noindent where $\mathbf{l_\alpha}$ are orthonormal spatial vectors; $\hat{P}_n$ is the operator of projection to the ground hyperfine level with the total angular momentum $F=n$; symbol $\hat\Sigma_{n,\alpha}$ denotes a Pauli matrix equivalent to a spin-\textit{n} particle; value $S_{[n],\alpha}$ is a component of spin polarization formed by $\mathrm{Rb}_{F=n}$ atoms; cropped density matrix $\hat\rho_n$ belongs to a subspace of $F=n$ hyperfine level, its rank is equal to $2n+1$ according to the number of Zeeman sublevels. Note, the cropped density matrix is averaged over alkali velocities.

As the vectors above have been defined, we come to a solution of the master equation~\eqref{master:equation} with satisfied conditions of non-adiabatic dynamics. The solution is density matrix dependent on time for any frequency $\Omega$ of the magnetic field defined by Eq.~\eqref{mag:field}. For the last period of the solution, we calculate spin polarization components via Eqs.~\eqref{S:n} --~\eqref{ro:n} for a wide range of frequencies $\Omega$. Finally, frequency dependence of two convolutions are presented in the Fig.~\ref{pic:freq:dep}, the latter demonstrates some peaks of spin polarization:
\begin{gather}
\mathcal{C}_1(\Omega)=\max\limits_{t\in T_{last}}\,\left|\mathbf{S_{[2]}}(t)\right|=\mathrm{sup}\,\left|\:\mathbf{S_{[2]}}\:\right|,\label{conv:1}\\[0.2cm]
\mathcal{C}_2(\Omega)=
\left|\:\frac{1}{T}\!\!\!
\int\limits_{t\in T_{last}}\!\!\!\!\!\mathbf{S_{[2]}}(t)\,\mathrm{d}t\:\right|=\left|\:\overline{\mathbf{S_{[2]}}}\:\right|.\label{conv:2}
\end{gather}

\noindent Regarding the selective scanning of alkali spin polarization (see Fig.~\ref{pic:levels}), we observe only $\mathbf{S_{[2]}}$ vector.

These convolutions give general information about spin dynamics. The first one describes the radius of the sphere that contains a trajectory of the vector $\mathbf{S_{[2]}}$ moving in the three-dimensional space of spin polarization components. The larger the $\mathcal{C}_1$, the more oriented alkali spins are. The second one describes alkali vapor magnetization, determined by the mean spin polarization over time. It can be measured by rotation of the linearly polarized scanning light, i.e. angle $\Delta\psi$ in the Fig.~\ref{pic:scheme} and Sec.~\ref{subsection:scanning:light}.

According to Fig.~\ref{pic:freq:dep}, the highest peak is the second from the right, it appears near the frequency $\Omega\approx 33.2\ kHz$. There are no additional peaks in the region above $50\ kHz$. Low frequencies are not considered in the work, since corresponding dynamics are close to adiabatic case there.

\subsection{Width of the highest peak}\label{second:peak}

Relaxation of a spin polarization occurs due to population mixing among the ground Zeeman sublevels, see the Table~\ref{table1}. This process is defined by the relaxation rate $\Gamma$, which determines the form of peaks.

In the Fig.~\ref{pic:freq:dep}, the second from the right (at $\approx33.2\ kHz$) peak of the convolution $\mathcal{C}_2$ is of interest. The high altitude of the peak is more important than its narrow width from the practical point of view: the less altitude, the lower signal-to-noise ratio. It is so since approximately only one alkali atom in a hundred creates spin polarization, hence detection of the effect may be challenging. Also, the $\mathcal{C}_2$ is more contrast compare to the convolution $\mathcal{C}_1$.


\begin{figure}[b]
\includegraphics[width=8cm]{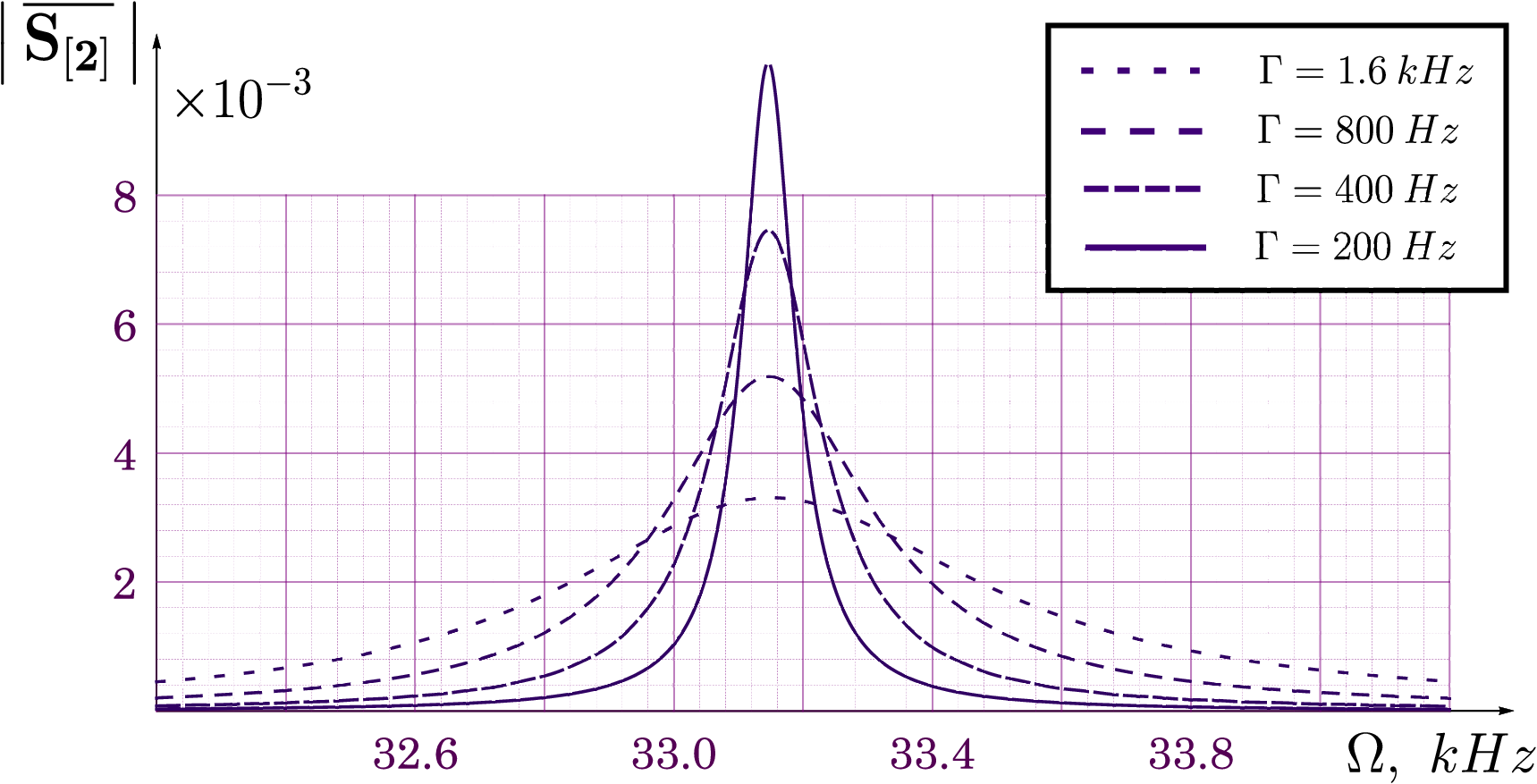}
\caption{\it \small The second from the right peak in the Fig.~\ref{pic:freq:dep} calculated for the different relaxation rates $\Gamma$. There is only convolution Eq.~\eqref{conv:2} in the picture}\label{pic:res}
\end{figure}

In Fig.~\ref{pic:res} one may observe dependence of its shape on different rates of $\Gamma$. Note, that the peak altitude and relaxation rate $\Gamma$ are not in inverse proportion. Therefore, a {\it the spin effect} behavior differs from a driven damped oscillator in condition of the resonance, when an amplitude of oscillations and the damping ratio are inverse proportional. Furthermore, the results are not well approximated by the Lorentz function as for a damped oscillator. Regarding the latter, we consider steep increase of spin polarization as non-adiabatic effect of spin dynamics, though its physical nature is different from the EPR.

\begin{figure}[t]
\includegraphics[width=8cm]{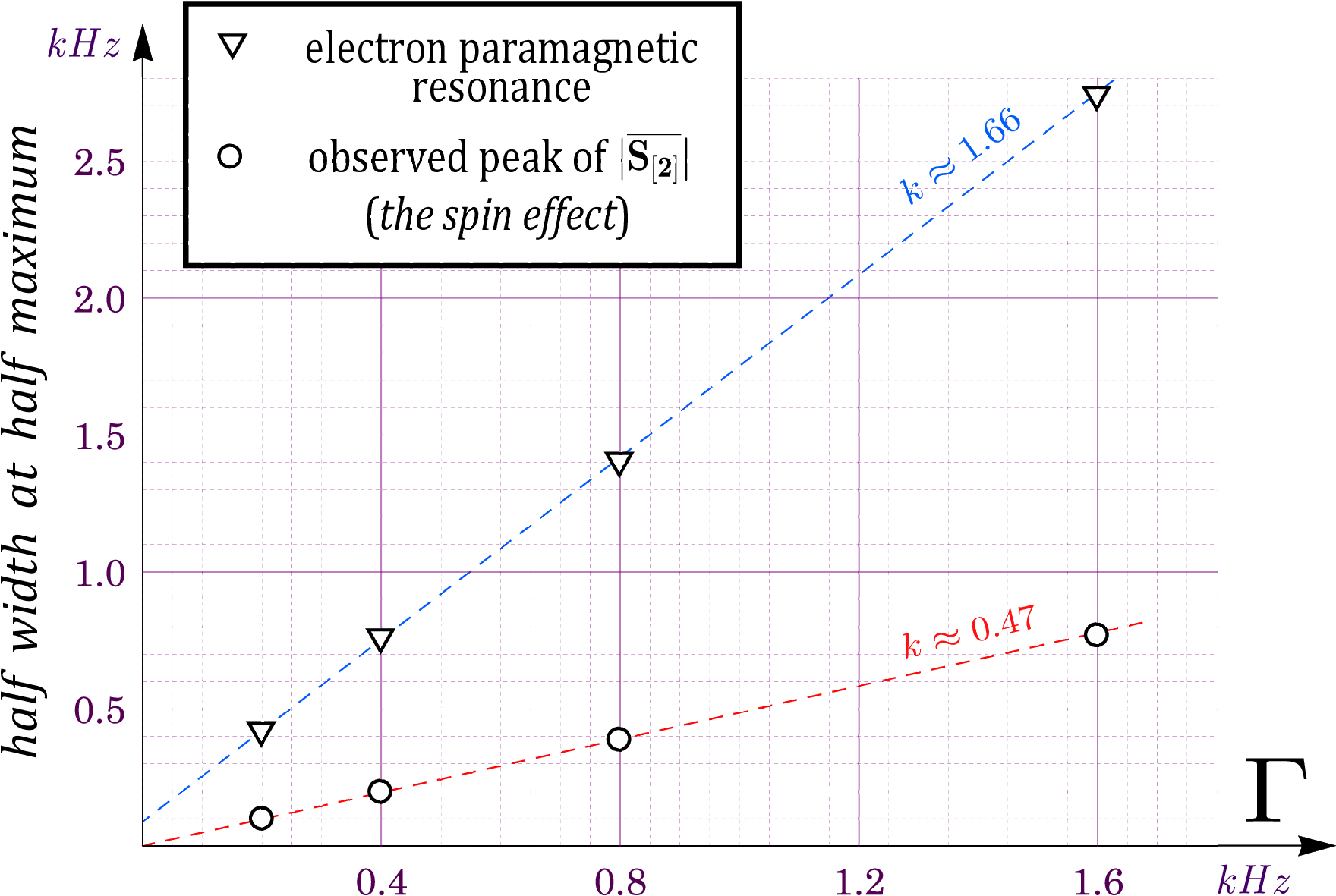}
\caption{\it \small Comparison of the peak's width in the Fig.~\ref{pic:res} with the width of the EPR calculated under the same conditions of pumping and relaxation. When the EPR was modeling, a magnetic field was defined by Eq.~\eqref{mag:epr} with $B_{dc}=10\ \mu T$ and $B_{ac}=10\  nT$. These parameters are chosen to match the frequency of the peak ($\approx33.2\  kHz$)}\label{pic:width}
\end{figure}

Another observable parameter is a width of the peak dependent on relaxation rate $\Gamma$, we specify it by the {\it half width at half maximum} (HWHM). We would like to compare the width of the peak dependence on relaxation rates $\Gamma$ with the corresponding dependence in case of the EPR under the same conditions of pumping and relaxation. The data for EPR is obtained by solving the master equation~\eqref{master:equation} implying magnetic field is defined as follows (instead of Eq.~\eqref{mag:field} for {\it the spin effect}):
\begin{equation}\label{mag:epr}
\mathbf{B}=B_{dc}\mathbf{l_z}+B_{ac}\mathbf{l_x}\cos\left(\Omega t\right),\qquad B_{dc} \gg B_{ac},
\end{equation}
\noindent where the constant magnetic field $B_{dc}$ determines the resonance frequency, and the alternating magnetic field with the amplitude $B_{ac}$ induces the resonance dynamics of alkali spin. In the Fig.~\ref{pic:width} one may observe calculated HWHMs for a given relaxation rates $\Gamma$ as in Fig.~\ref{pic:res}. As a result, we estimate that the HWHM of the observed peak for {\it the spin effect} is $3.5$ times less than the one of the EPR. Note, that HWHM and relaxation rate are proportional in both cases.

\subsection{Spin dynamics near peak polarization}\label{second:peak:trajectory}

\begin{figure*}[t]
\includegraphics[width=17cm]{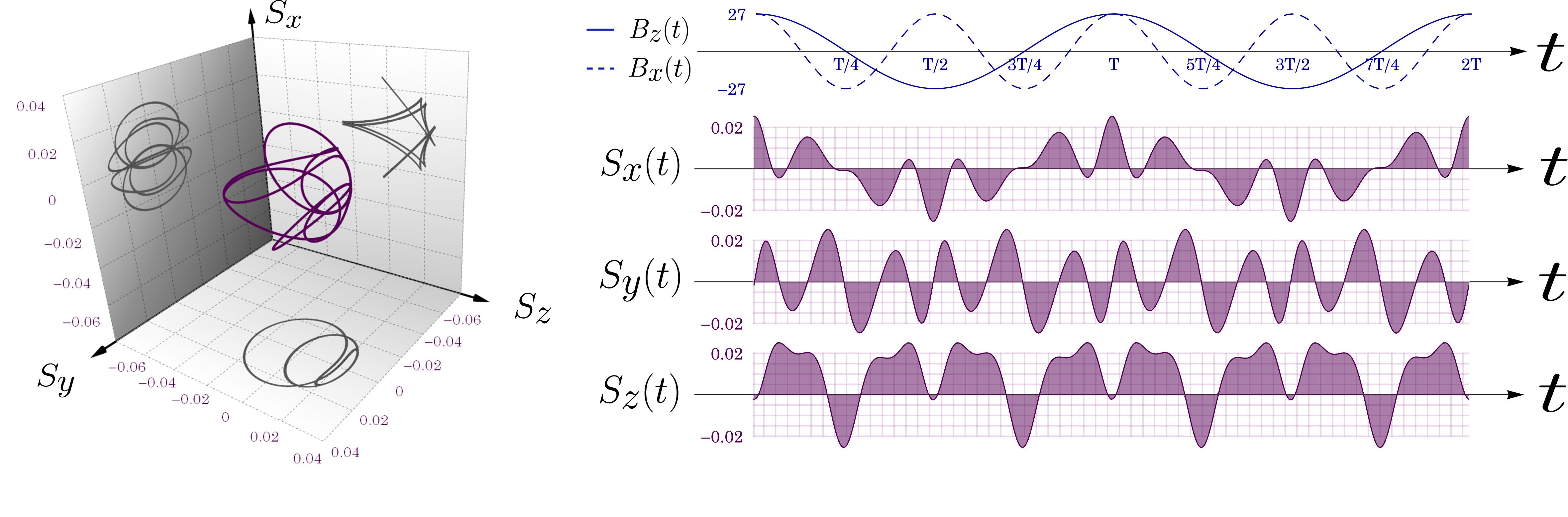}
\caption{\it \small The purple parametric 3d-curve drawn by three temporal dependencies of the spin polarization components $S_{[2],\alpha}$, where $\alpha\in\{x,y,z\}$. There is a steady dynamics of spin polarization shown in the picture. Blue curves demonstrate two components of the periodic magnetic field, defined by Eq.~\eqref{mag:field}. Note, the index $[2]$ in $S_{[2],\alpha}$ is dismissed for the sake of visual simplicity. The purple curves are calculated for the frequency $\Omega\approx 33.2\ kHz$, which matches the maximum of the peak in the Fig.~\ref{pic:freq:dep}}\label{pic:trajectory}
\end{figure*}

Further we explore the steady non-adiabatic dynamics of the spin polarization created by atoms $\mathrm{Rb}_{F=2}$, when the frequency of the magnetic field, defined by Eq.~\eqref{mag:field}, maximizes $\left|\overline{\mathbf{S_{[2]}}}\right|$, i.e. $\Omega\approx 33.2\ kHz$, see the Fig.~\ref{pic:freq:dep}. We consider the three components of the spin polarization, which can be measured by scanning light: $S_{[2],x}(t)$, $S_{[2],y}(t)$, and $S_{[2],z}(t)$. The trajectory of the vector $\mathbf{S_{[2]}}(t)$ in three-dimension space of spin polarization components and its parametric representation are plotted in the Fig.~\ref{pic:trajectory}. It is mirror symmetric about two planes: $S_xS_z$ and $S_yS_z$. A little deviation from the mirror symmetry about $S_xS_z$~the plane occurs due to location of the magnetic field in the $XZ$~plane, see Fig.~\ref{pic:scheme}. At the same time, the trajectory is not mirror symmetric about the $S_xS_y$~plane. We assume that the axis of the general shift of the spin polarization is correlated to the direction of pump light.


The average radius and the Z-shift of the trajectory is described by two convolutions $\mathcal{C}_1$ and $\mathcal{C}_2$, defined in the Eqs.~\eqref{conv:1} and~\eqref{conv:2}. So alteration of these trajectory characteristics may be estimated by the observation of the latter convolutions dependence on frequency $\Omega$, shown in Fig.~\ref{pic:freq:dep}. If the frequency deviates from the value at the point of maximum, the trajectory collapses to the axis-origin and reshapes to isotropic tangle clew. Reshaping of the trajectory explains the higher rate of the collapse of $\mathcal{C}_2$ convolution compare to the one of $\mathcal{C}_1$ convolution. Note, the spin polarization trajectories are different for each of four peaks shown in the Fig.~\ref{pic:freq:dep}. Despite we provide the figure for only one, all other trajectories exhibit similar properties: two mirror symmetries and “protuberant” along the path of pump light propagation. To observe correlations, time dependencies $S_\alpha(t)$ in the Fig.~\ref{pic:trajectory} are plotted together with the magnetic field dynamics. Surprisingly, non-adiabatic dynamics of spin polarization involves strong higher harmonics, which are absent in the dynamics of the magnetic field.

\section{Analysis}\label{sec_analysis}
We investigate the nature of steep increase of an alkali spin polarization under non-adiabatic dynamics in presence of the strong alternating magnetic field. As discussed in the previous section, {\it the spin effect} appears to be not a resonance. We claim that {\it the spin effect} is a phenomenon of unpredictable periodic spin dynamics, that lasts while spin state decays. It occurs only in the non-adiabatic case. Then, the main condition of {\it the spin effect} to be observed is a synchronous spin motion inside an alkali vapor. If atomic spins are not synchronized, the ensemble of atoms tends to a mixed spin state with equilibrium distribution, and the spin polarization is vanished. Note, that observed effect does not require the same spin dynamics for all atoms.

In case of EPR, one component of the vapor magnetization approaches to a saturated state and then does not vary; this component is often called {\it longitudinal}. At the same time, both {\it transverse} components rapidly oscillate, but in most cases its magnitudes are small in comparison to the {\it longitudinal} component. Thus, EPR cannot be interpreted as a non-adiabatic process.

In order to explain the synchronous spin motion, let us describe an alkali atom $\mathrm{Rb}_{L=2}$, that is moving as free particle between two time points: the moment of initial spin state creation due to pump photon absorption, and the moment of spin state decay due to collision with the cell walls or with another alkali atom. Time interval of the free path of motion is estimated by inverse of relaxation rate $\Gamma$. As the first order approximation, we can use the Pauli equation to model a free alkali atom in an alternating magnetic field:
\begin{gather}
i\hbar\dot{\varphi}=\gamma\hbar\left(\hat{\sigma}\cdot\mathbf{B}(t)\right)\varphi,\label{Pauli}\\
\hat{\sigma}=\frac{1}{2}\left(\hat{\sigma}_x \mathbf{l_x}+\hat{\sigma}_y \mathbf{l_y}+\hat{\sigma}_z \mathbf{l_z}\right),\\
\varphi(t)=\hat{U}(t,t_0)\varphi_0,
\end{gather}
\noindent where $\varphi$ is the spinor, $\gamma$ is the gyromagnetic ratio, $\hat{\sigma}_\alpha$ are Pauli matrices, where $\alpha\in\{x,y,z\}$, $\varphi_0$ is the spinor state at the initial time $t_0$, $\hat{U}(t,t_0)$ is the unitary time-evolution operator.

\begin{table}[h]
\caption{\centering \it \small The highest four frequencies, that lead to periodic solution of the Pauli equation~\eqref{Pauli}, and corresponding solution properties}\label{table:loops}
\begin{ruledtabular}
\begin{tabular}{ccccc}
        Frequency & Initial spin & \multicolumn{3}{c}{Spin components}\\
        from the set &      direction & \multicolumn{3}{c}{averaged over a period ($\times 10^3$)}\\
        \multirow{2}{*}{$\Omega_{\mathcal{A}}/\gamma B_0$} &
        \multirow{2}{*}{$\langle\varphi_0|\hat\sigma|\varphi_0\rangle$} &
        \multirow{2}{*}{$\overline{\langle\varphi|\hat\sigma_x|\varphi\rangle}$} & \multirow{2}{*}{$\overline{\langle\varphi|\hat\sigma_y|\varphi\rangle}$}
        & \multirow{2}{*}{$\overline{\langle\varphi|\hat\sigma_z|\varphi\rangle}$}
        \\
        \\
        \colrule
        \\
        \multirow{3}{*}{$0.099$}
        & $=\mathbf{l_x}/2 $ & -48.4 & 0 & -26.4 \\
        & $=\mathbf{l_y}/2$ & 0 & -5.8 & 0 \\
        & $=\mathbf{l_z}/2$ & -22.6 & 0 & 56.3 \\
	\\
        \multirow{3}{*}{$0.126$}
        & $=\mathbf{l_x}/2 $ & 1.5 & 0 & -56.9 \\
        & $=\mathbf{l_y}/2$ & 0 & 31.7 & 0 \\
        & $=\mathbf{l_z}/2$ & 1.6 & 0 & 57.5 \\
        \\
        \multirow{3}{*}{$0.175$}
        & $=\mathbf{l_x}/2 $ & -4.7 & 0 & 196.8 \\
        & $=\mathbf{l_y}/2$ & 0 & -35.6 & 0 \\
        & $=\mathbf{l_z}/2$ & -52.5 & 0 & -17.6 \\
        \\
        \multirow{3}{*}{$0.259$}
        & $=\mathbf{l_x}/2 $ & -44.3 & 0 & 5.5 \\
        & $=\mathbf{l_y}/2$ & 0 & -82.2 & 0 \\
        & $=\mathbf{l_z}/2$ & 4.2 & 0 & 58.4 \\
        \\
\end{tabular}
\end{ruledtabular}
\end{table}

The Pauli equation~\eqref{Pauli} with the alternating magnetic field $\mathbf{B}(t)$, defined by Eq.~\eqref{mag:field}, generates a set $\mathcal{A}$ of frequencies, which lead to a periodic solution independent of the spinor state at the initial time:
\begin{equation}
\forall \varphi(t):\quad\varphi(t+2\pi/\Omega_\mathcal{A})=\varphi(t),
\end{equation}
where $\Omega_\mathcal{A}\in\mathcal{A}$. The set $\mathcal{A}$ is discrete and upper-bounded by a maximal frequency. A sequence of the highest frequencies in the set $\mathcal{A}$ corresponds to coordinates of the peaks in the Fig.~\ref{pic:freq:dep}. Note, that if a frequency from the absolute complement of $\mathcal{A}$ is substituted into the Pauli equation, the solution can be periodic only for the certain initial state.

In the Table~\ref{table:loops} we present the four highest frequencies from the set $\mathcal{A}$ divided by $\gamma B_0$ and corresponding properties of the Pauli equation solutions. Although, periodicity does not depend on an initial state for the frequency $\Omega_\mathcal{A}$, the solution of the Pauli equation~\eqref{Pauli} does. To describe a behavior of the spin dynamics, we define spin components averaged over a period. The latter corresponds to the second convolution expressed by Eq.~\eqref{conv:2}. It is evaluated for a given three (orthogonal) initial states and four different frequencies $\Omega_\mathcal{A}$.

Data in the Table~\ref{table:loops} shows us, that maximal magnetization occurs when frequency $\Omega$ is equal to the second highest $\Omega_\mathcal{A}$ ($\approx 33.2\ kHz$ and corresponds to $\Omega_\mathcal{A}/\gamma B_0\approx 0.175$). The averaged spin polarization is quasi-collinear with the Z-axis, which matches with the path of the pump light propagation.

Based on the arguments above, we propose the following statements that explain {\it the studied spin effect}:
\begin{enumerate}
    \item Permanent relaxation and pump of the alkali spin polarization leads to jumps of atomic spin between different quantum states;
   \item If the dynamics is non-adiabatic and periodic, alkali spins can synchronously move starting from a random the initial spin state;
   \item Condition of the synchronous moving is the certain frequency/amplitude rate of an alternating magnetic field, which is determined by corresponding temporal profile;
   \item The effect of synchronized and repetitive motion of the spins appears as steep increase of an averaged spin polarization over a period.
\end{enumerate}

\section{Conclusion}\label{sec_conclusion}

We have considered a non-adiabatic spin polarization dynamics, which occurs in alkali vapor under optical spin orientation and in the presence of a strong alternating magnetic field. By solving the master equation for a density matrix, we have revealed the effect that resembles a resonance. If the frequency of a periodic magnetic field without a constant component is equal to a certain value, an alkali spin polarization steeply increases. However, with a certainty we claim, that it is not an electron spin resonance, as it is revealed by the analysis of the peaks that may be observed in the dependence of the spin polarization on the magnetic field frequency. We interpret the effect as a fundamental property of non-adiabatic spin polarization dynamics in presence of the external magnetic field, when spins that belong to an alkali ensemble move synchronously only for a certain frequency. Mathematically, it can be explained by the existence of Pauli equation solutions with periodic behavior, if periodic magnetic field has the certain profile.

The important result from the practical point of view is that the width of one of the found peaks is significantly narrower than the width of EPR induced in the same medium with similar conditions, see Fig.~\ref{pic:width}. Therefore, the effect may find its place in the field of precise sensing.

\section*{Funding}
This work was financially supported by Russian Ministry of Education (Grant No. 2019-0903)

\begin{acknowledgments}
We thank our colleagues A.D. Kiselev and G.P. Miroshnichenko for fruitful discussions during the research.
\end{acknowledgments}

\nocite{*}

\bibliography{apssamp_red}

\end{document}